\documentclass[sigconf]{acmart}
\usepackage{graphicx}
\AtBeginDocument{%
  \providecommand\BibTeX{{%
    \normalfont B\kern-0.5em{\scshape i\kern-0.25em b}\kern-0.8em\TeX}}}

\copyrightyear{2021}
\acmYear{2021}
\setcopyright{acmlicensed}\acmConference[CEA '21]{Proceedings of the 13th International Workshop on Multimedia for Cooking and Eating Activities}{August 21--24, 2021}{Taipei, Taiwan}
\acmBooktitle{Proceedings of the 13th International Workshop on Multimedia for Cooking and Eating Activities (CEA '21), August 21--24, 2021, Taipei, Taiwan}
\acmPrice{15.00}
\acmDOI{10.1145/3463947.3469235}
\acmISBN{978-1-4503-8532-9/21/08}

\begin{document}

\title{World Food Atlas Project}

\author{Ali Rostami}
\email{rostami1@uci.edu}
\affiliation{%
  \institution{The University of Irvine}
  \country{USA}
}

\author{Zhouhang Xie}
\email{zhouhanx@uci.edu}
\affiliation{%
  \institution{The University of Irvine}
  \country{USA}
}

\author{Akihisa Ishino}
\email{ishino@hal.t.u-tokyo.ac.jp}
\affiliation{%
  \institution{The University of Tokyo}
  \country{Japan}}

\author{Yoko Yamakata}
\email{yamakata@mi.u-tokyo.ac.jp}
\affiliation{%
  \institution{The University of Tokyo}
  \country{Japan}}

\author{Kiyoharu Aizawa}
\email{aizawa@hal.t.u-tokyo.ac.jp}
\affiliation{%
  \institution{The University of Tokyo}
  \country{Japan}}

\author{Ramesh Jain}
\email{jain@ics.uci.edu}
\affiliation{%
  \institution{The University of Irvine}
  \country{USA}
}

\begin{abstract}
A coronavirus pandemic is forcing people to be "at home" all over the world.
In a life of hardly ever going out, we would have realized how the food we eat affects our bodies.
What can we do to know our food more and control it better?
To give us a clue, we are trying to build a World Food Atlas (WFA) that collects all the knowledge about food in the world.
In this paper, we present two of our trials.
The first is the Food Knowledge Graph (FKG), which is a graphical representation of knowledge about food and ingredient relationships derived from recipes and food nutrition data.
The second is the FoodLog Athl and the RecipeLog that are applications for collecting people's detailed records about food habit.
We also discuss several problems that we try to solve to build the WFA by integrating these two ideas.
\end{abstract}



\maketitle

\section{Introduction} 

Food determines human quality of life. Interestingly, food production determines the quality of planetary environment.  As we know, food provides the energy and nutrients essential for health and is also a major source of personal enjoyment and social fabric in human society.   To make an interesting situation become a major challenge in society, pleasures of eating conflict with the optimal nutritional needs of the person’s physiological well-being, resulting in the substantial increase in diseases.   In fact, there is an interesting challenge similar to the famous ‘mind-body problem’ related to food: ‘enjoyment-body problem’.  What I enjoy eating is not necessarily what my body wants me to eat.  Can we satisfy both me and my body? 

Technology may help in addressing this enjoyment-body problem by formulating this as a recommendation problem \cite{Rostami2020PersonalModel}, \cite{Min2020FoodChallenges}.  A major challenge in addressing this problem is to create a modern electronic World Food Atlas (WFA). This WFA will contain information related to all dishes and food items that are accessible to a person at a location anywhere in the world.  Given the variations possible in consumable food and variety of ingredients and recipes available at any point, it is a daunting challenge to build a WFA.  But it is an essential problem for not only enhancing the quality of life of each individual, but for the survival of the planet also.  We have taken an early step towards this ambitious goal.  Our goal is to start the process of building a global community of scientists, technologists, and food related experts to build the WFA.  This paper describes an early step towards that.

\begin{figure}[!ht]
  \centering
  
  \includegraphics[width=0.7\linewidth]%
    {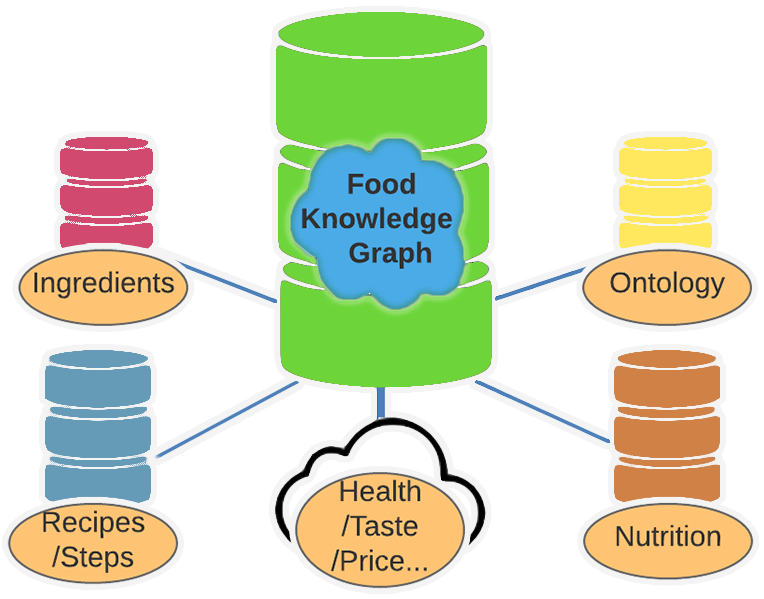}
    \caption{Food Knowledge Graph aggregating and unifying food-related information from multiple offline and online sources.}
    \label{fig:FKGflow}
\end{figure}

\begin{figure*}[h]
  \includegraphics[width=0.8\textwidth]{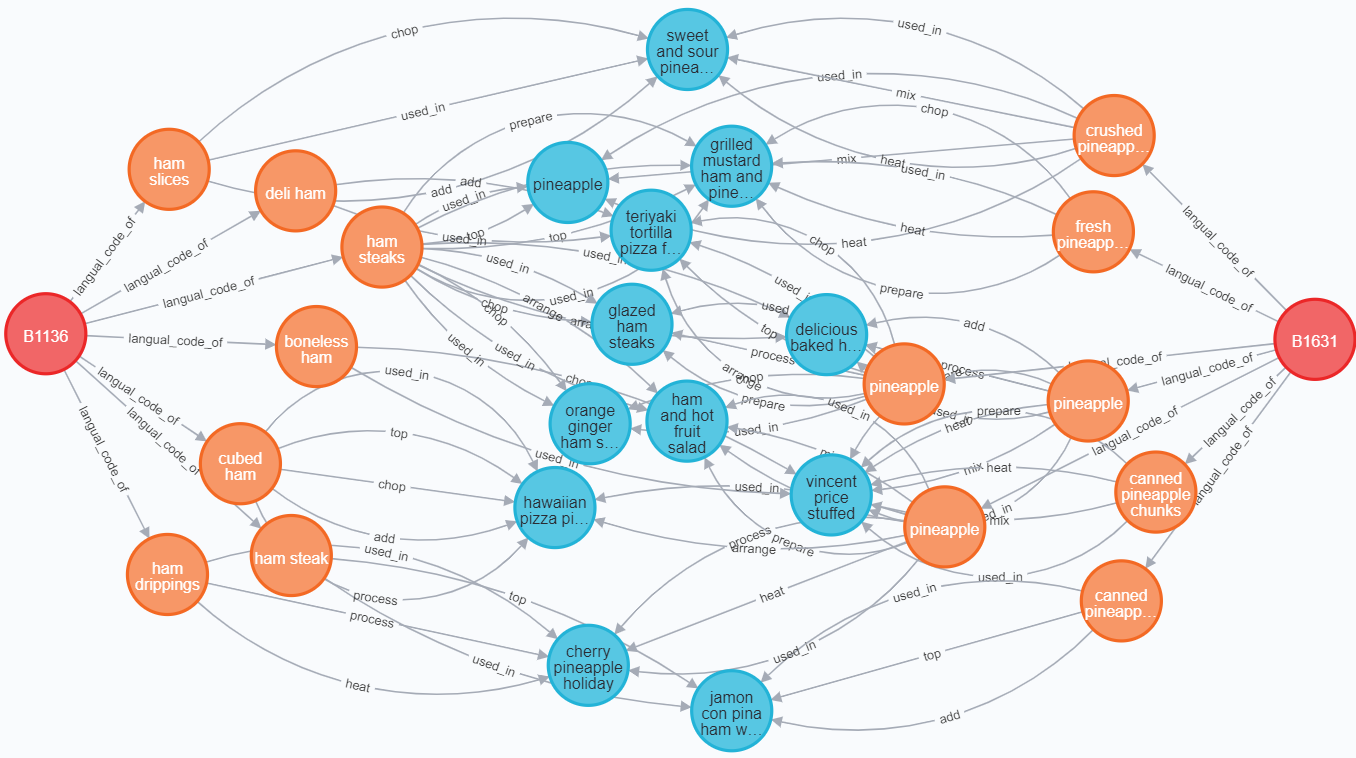}
  \caption{Food Knowledge Graph representation in the Neo4j environment. B1136 is the general pork node and B1631 is the general pineapple node. This image shows the result of the query looking for food items which have both pork and pineapple as ingredients.}
  \label{fig:PFMSystemsOverview}
\end{figure*}
\section{Food Knowledge Graph} 

The mission of WFA is beyond its technology but the core part of the WFA is the Food Knowledge Graph (FKG). FKG is basically the collection of interlinked descriptive information about different aspects of food. FKG contains different dimensions of connectivity and different node types. The ingredient nodes are interconnected by a semantically hierarchical structure obtained from the food ontology database \cite{Dooley2018FoodIntegration} which provides a cue about the similarity between different ingredients and acts as the backbone of the formal semantics of the FKG. Some nodes in the ingredient category are actual ingredient nodes such as cow's milk or chicken, and some other ingredient nodes are abstract nodes which represent a class of ingredients such as dairy or meat. The hierarchical connections between the ingredient nodes makes sure that ingredients within the same category always have a shorter path to each other than to nodes outside of the category. For example cow's milk should have a shorter distance to Greek yogurt than to chicken since cow's milk and Greek yogurt are both a subset of the dairy node. More complex nodes are the recipe nodes, in another dimension of connectivity, the recipe nodes are connected to the corresponding ingredient nodes. The edge connecting a recipe node to its corresponding ingredients contains information about the portion and also information about the step in which the ingredient is involved in, such as frying or mixing. Entity nodes must be able to contain geographical information such as geographical availability, popularity and origin of dishes and some ingredients. Furthermore the detailed nutritional content of ingredients are associate to ingredient nodes which enables the derivation of the nutritional contents of the dish nodes since they are connected to the ingredient nodes. The entity descriptions contribute to one another by providing context for further interpretations which is the main promise of the FKG. In order to create the FKG data from multiple sources need to be brought together to form a foundation which represents the food knowledge as shown in Figure \ref{fig:FKGflow}. Once the FKG is designed with some initial nodes allowing both people and computers to process these formal semantics an efficient and unambiguous manner, expansion standards can enable public contribution from parties with various backgrounds in a semi-supervised manner. However creating the initial backbone of the FKG is in facet a challenging task since it involves problems which are currently hot research topics in different fields such as multimedia processing, knowledge graph formation and optimization, food entity resolution, query refinement and natural language processing. Figure \ref{fig:PFMSystemsOverview} shows a sub-graph of the FKG which is the result which was generated for a query asking for all the food items which contain both pork and pineapple. This section is currently being implemented by the Food Computing team at the University of California.

\section{Detailed food record collection} 

The other group in this project, the University of Tokyo's Aizawa-Yamakata Lab, is developing FoodLog Athl~\cite{Aizawa2019FoodLog:Application}, which creates food records semi-automatically by recognizing images of food, and RecipeLog, a smartphone application that allows chefs to easily record their own recipes with the assistance of AI. By combining both of these services, it will be possible to accurately record food history in the home and monitor nutritional intake. The aim of this proposal is to help people understand their own food and acquire the habit of managing their dietary habits appropriately.

\subsection{Food Logging App: FoodLog Athl}
\begin{figure}[ht]
  \centering
\begin{tabular}{cc}
       \includegraphics[width=0.45\linewidth]%
    {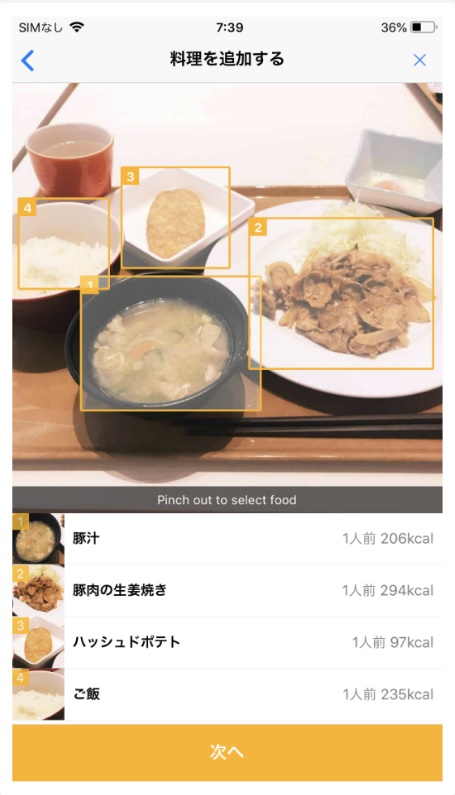}
    &  
       \includegraphics[width=0.45\linewidth]%
    {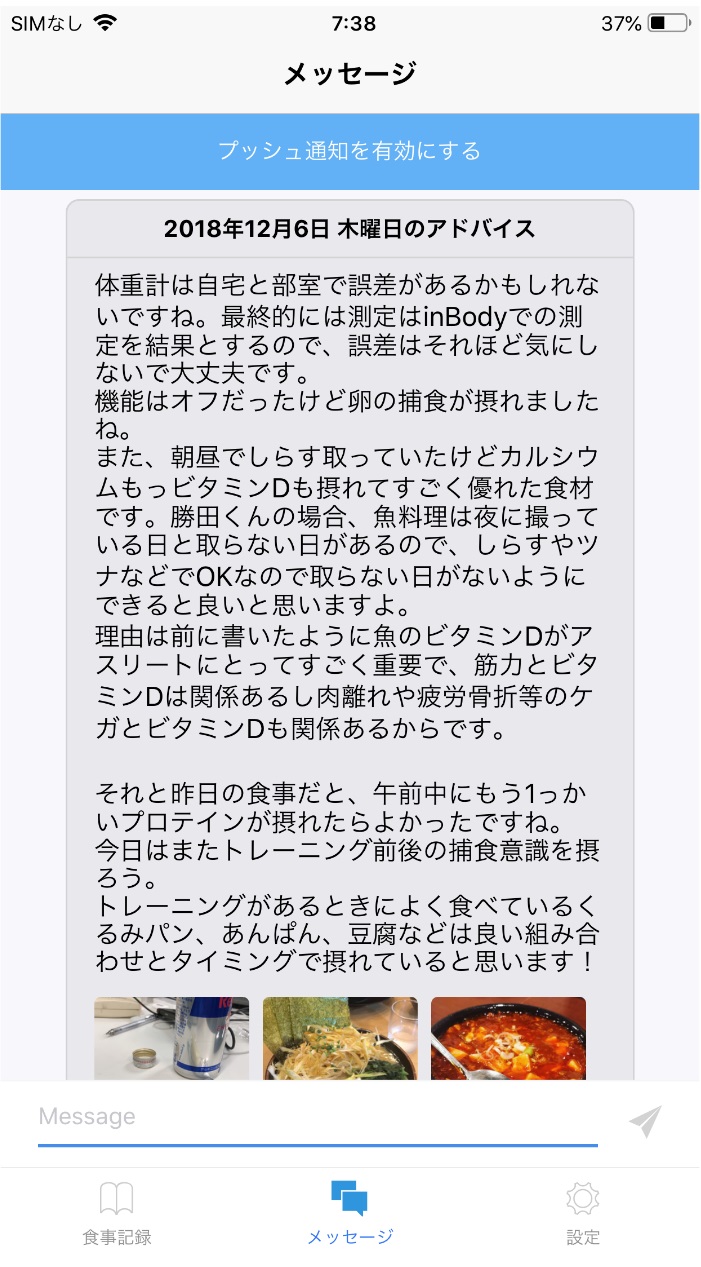}
    \\
     (a) Construct food record & (b) Chat with nutritionist \\ [-0.2em]
     using image recognition. & to have their advices.
\end{tabular}
    \caption{FoodLog Athl creates food records semi-automatically by recognizing images of food. It has the function of communicating with a nutritionist via chat.}
    \label{fig:FoodLog}
\end{figure}

\begin{figure}[htb]
  \centering
\begin{tabular}{cc}
       \includegraphics[width=0.33\linewidth]%
    {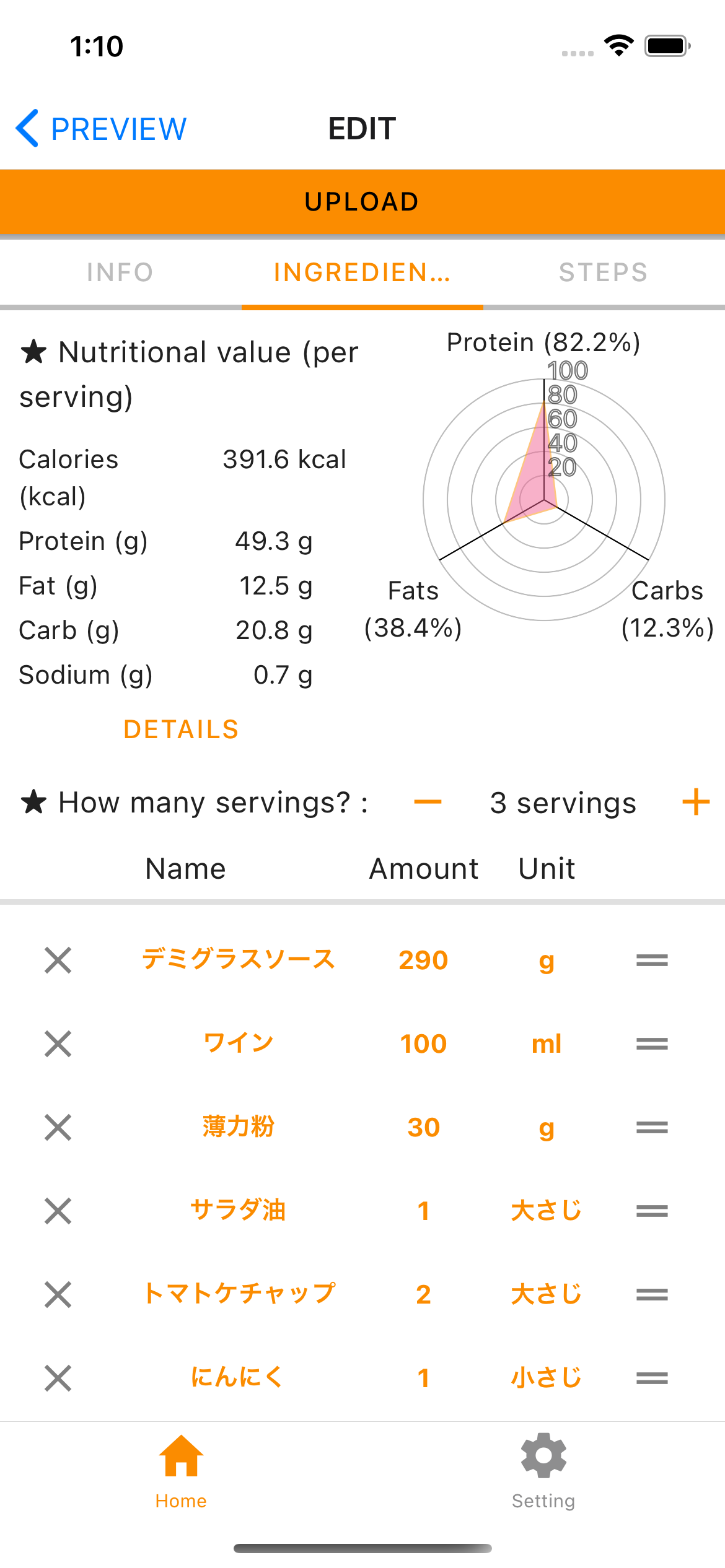}
    &  
       \includegraphics[width=0.33\linewidth]%
    {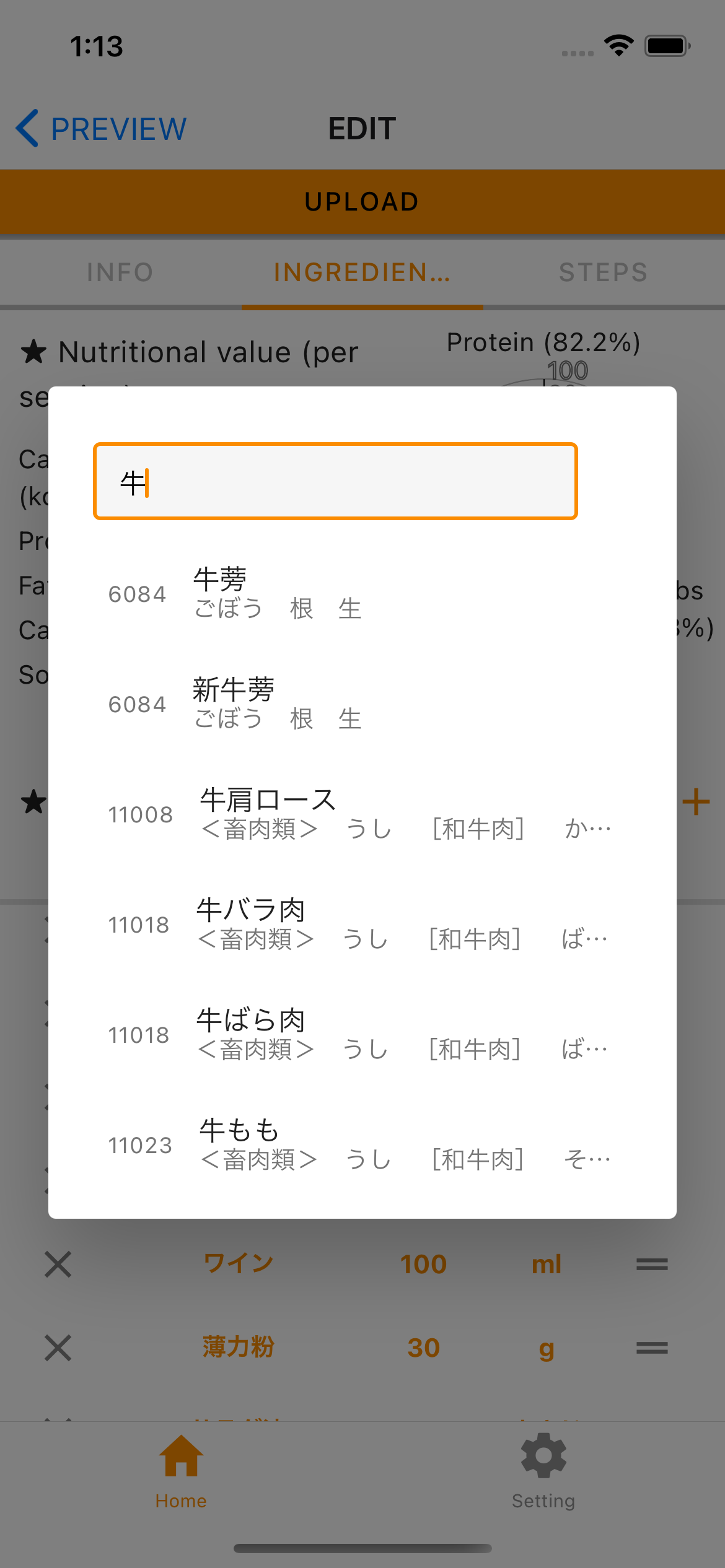}
    \\
     (a) Input ingredients & (b) Ingredient search
\end{tabular}
    \caption{The RecipeLog provides the function of authoring recipes. (a) Once the ingredients are entered, their nutritional value is visualized in a radar chart. (b) The ingredients are searched from the "standard tables of food composition" provided by the Ministry of Education, Culture, Sports, Science and Technology.}
    \label{fig:RecipeLog}
\end{figure}

In 2019, we released FoodLog Athl (Figure \ref{fig:FoodLog}, \url{http://www.foodlog-athl.org/}).
Using deep learning technology, it recognizes multiple dishes and suggest the users to record the food names. Although the number of initial recognizable classes of food items is limited to about 500, the number increases by learning personal records of the users. In terms of the food items with nutrition data, 2000 generic foods and more than 100,000 items of products or restaurant chains are available. 
This tool is designed not only for taking user's food record, but also to help dietitians take care of their patients. We also provide tools for dietitians to monitor and communicate with their patients' diets.
Users' meals, especially those eaten at home, are based on specific recipes that no one knows.
Differences, additions, or omissions of ingredients can result in very different nutritional values. By linking to the RecipeLog (Figure \ref{fig:RecipeLog}), dietitians can see an accurate nutritional record based on their patients' recipes.

\subsection{Recipe Authoring App: RecipeLog}

Dietary habits are a major factor in causing lifestyle-related diseases, and when doctors and nutritionists intervene, guidance is provided based on records of the names of dishes that were eaten each day.
However, dishes prepared at home are often arranged by each family for reasons such as cost, taste, and cooking time, and the nutritional data registered in the database cannot be relied upon.
On the other hand, the person who prepared the dish knows the ingredients used in detail, and if the recipe is registered, detailed nutritional data can be obtained.

One of the challenges of the RecipeLog is that for many users, writing a recipe is a chore.
Home chefs may change some of ingredients depending on seasons often, and they may not want to write a new recipe every time.
Therefore, the recipe log will provide a function to search for the recipe in the database that is closest to the one the user wants to make, and create a new recipe by editing it based on that.
Users can create a new recipe simply by rewriting a recipe that contains the basic method of making the dish, or a recipe that they have written in the past.
Recipes created in this way not only reflect the actual recipes of each family in detail, but also clarify the differences between recipes by remembering where they have been rewritten, just like a version control system, such as "Git".
This characteristic is what makes our dataset different from other datasets of recipes commonly used in the multimedia field, such as Recipe 1M+\cite{Marin2021Recipe1M+:Images}.

Cooking procedures are also essential for rigorous nutritional assessment, but writing cooking instructions must be a hassle for users.
Therefore, the RecipeLog predicts instruction candidates based on the name of the dish and the list of ingredients, and presents them to the user.
The academic challenge is to obtain the correct workflow with as little user editing as possible.
In fact, several methods have been proposed to generate recipes from pieces of information such as food images, titles, and ingredient lists.
Salvador et al. proposed a method to generate recipes from food images\cite{Salvador2019InverseImages}. Also Kiddon et al. proposed a method to generate a recipe from title and ingredients\cite{kiddon-etal-2016-globally}. A method to generate personalized recipes are also proposed by Majumder et al\cite{majumder-etal-2019-generating}. 
This kind of research could be used to assist users in writing recipes.




\section{discussion} 

Recipes are the backbone of the FKG and in order to make the FKG well structured and scalable, standardized recipes are essential.
Although the flexibility of expression in the RecipeLog is limited compared to general recipes written in free text, it can provide such recipes in which most of the ingredients are linked to the food nutrition facts and there is less fluctuation in the written instructions.
In this sense, we believe that the integration of FKG and RecipeLog into a single service will bring a new innovation to the world of food.
Here, we will discuss the problems we need to tackle and the proposed solutions.

\noindent{\bf How to manage the stress of controlling diet?} 

A considerable effort behind this work is highly motivated by the importance of emerging personal food modeling techniques used in modern diet management platforms. A Personal Food Model (PFM) \cite{Rostami2020PersonalModel} uses different life log data streams from the user as input and models the food related characteristics of the user which will be used to guide users towards their specific goals based on their unique personal preferences and reactions.  Recent work in PFM such as \cite{Pandey2021EventModelling} show that the combination of the food log history \cite{Rostami2020MultimediaLogger} with the context vector can be used to estimate a preferred food region for future contexts. However this kind of approach is currently very limited without the existent of FKG since only the FKG can provide truly connected data lake about different aspects of food to define a complex space in which certain foods lie within. Having interlinked information about food in form of a knowledge graph can provide a subset of the graph for queries searching for foods with certain conditions for various applications and diet management platforms can be the first users of it. 

\noindent{\bf How to overcome the differences in food culture?} 

The initial step starts with understanding what the differences are. What kind of food items are more popular in certain parts of the world and what food items are avoided by people in another region. The next step would be understanding what are the causal factors behind these differences. The causal aspect of the cultural food differences can be complex, it can be due to availability of certain food items in certain regions, it could be the difference between economy and average income, could be genetics or cultural and religious beliefs. But whatever it is, it is important to study and understand these differences so that people with different cultures understand each other better as the same body of people all around the world. This kind of information will be available with the combination of FKG and food logging platforms. 
Concretely, recipes written in different languages need to be standardized into FKG-compatible processes and entities, so that they could be analyzed in a unified framework. For the standardization process to be scalable and modular, there need to be language-specific pipelines that perform the recipe standardization process using natural language processing (NLP) methods, so that linking recipes to the FKG would only require aligning information from different languages on the level of entities and processes.


\noindent{\bf How to overcome the hassle of taking food records?} 

RecipeLog and FoodLog can support the precise food recording.

As mentioned above, what the RecipeLog wants to record is not an ordinary recipe, but only the work-flow, such as what ingredients were cooked and how they were processed.
Therefore, the expression space is extremely limited, and AI can provide powerful writing support.
For example, if you are making lasagna, you can predict that the onions will be finely chopped without any explanation.
You might explain such a process of peeling the onions if you are teaching someone how to cook, but it is not necessary for your records.
RecipeLog presents the user with a choice of cooking processes that can be derived probabilistically from the name of the dish and ingredients, and the user only has to select the appropriate procedure among them.
A lazy user may accept a procedure proposed by the system without correcting it, even though it is actually wrong.
One of the major challenges of the RecipeLog is how to accurately absorb the recipes of dishes that users truly cook at home by incorporating such user behavior.

The meal recording itself is completed by simply taking a picture of the meal using the service of FoodLog Athl.
FoodLog Athl already has a large number of active users and a proven track record.


\section{conclusion} %

This paper introduces two trials as a first step toward building the World Food Atlas.
The first is the Food Knowledge Graph (FKG), which is a graphical representation of knowledge about food and ingredient relationships derived from recipes and food nutrition data.
The second is the FoodLog Athl and the RecipeLog, mobile applications for collecting people's detailed records about food.
Although these studies have so far proceeded independently, by sharing multimodal data, they will solve some of the problems mentioned in the discussion section.
Food culture differs from region to region.
The goal is to develop the WFA globally so that people all over the world can have healthier and more enjoyable diets.

\section{Acknowledgement}
This paper and the research behind it would not have been possible without the exceptional support of Prof. Amarnath Gupta and Brian W. Stack as they had key role in creating the initial version of the Food Knowledge Graph.

\bibliographystyle{ACM-Reference-Format}
\bibliography{references2}
\end{document}